# Slow electron elastic scattering by a target represented by different zero-range potentials

A. S. Baltenkov[1] and I. Woiciechowski[2]

[1]*Arifov Institute of Ion-Plasma and Laser Technologies, 100125, Tashkent, Uzbekistan*
[2]*Alderson Broaddus University, 101 College Hill Drive, WV 26416, Philippi, USA*

**Abstract:** The general formulas to calculate the phase shifts of wave function of a particle scattering on a target formed by a pair of non-identical zero-range potentials are derived. The elastic scattering cross sections of slow electrons by $C_2$ and CH molecules, which are modeled by pairs of different zero-range potentials, are calculated. It is shown that at asymptotically great distances from the target the continuum wave function of particle is presented as an expansion in a set of other than spherical $Y_{lm}(\mathbf{r})$, orthonormal functions $Z_{\lambda}(\mathbf{r})$. General formulas for these functions are obtained. The special features of the S-matrix method for the case of arbitrary non-spherical potentials are discussed.

## Introduction

In [1,2], the method of partial waves was generalized for nonspherical scatterers. It was shown that for the case when the scatterer can be represented as a superposition of *N* potentials of zero range the exact solution of the scattering problem reduces to an algebraic problem of the inversion of some matrix of order *N*.

The approach developed in [1,2] was implemented in the article [3]. The general formulas for the phase shifts of multiple elastic s-scattering of a slow particle on targets consisting of two, three, and four identical zero-range centers equally distant from each other have been derived. It is interesting to follow how these formulas transform in the case when the zero-range potentials forming the target are not identical. In this paper, we focus on the study of two-center targets formed by a pair of different zero-range potentials.

Brueckner [4] exactly solved the problem of multiple scattering of a particle with the wave vector *k* by two identical fixed centers of zero radius. According to the Huygens-Fresnel principle, the scattering wave function was written in [4] as a combination of the plane wave and two spherical s-waves generated by the scattering centers from the points $\mathbf{R}_1$ and $\mathbf{R}_2$

$$\psi_{\mathbf{k}}^{+}(\mathbf{r}) = e^{i\mathbf{k}\cdot\mathbf{r}} + D_1(\mathbf{k})\frac{e^{ik|\mathbf{r}-\mathbf{R}_1|}}{|\mathbf{r}-\mathbf{R}_1|} + D_2(\mathbf{k})\frac{e^{ik|\mathbf{r}-\mathbf{R}_2|}}{|\mathbf{r}-\mathbf{R}_2|}. \tag{0}$$

Throughout this article, we use the atomic system of units. The coefficients at these spherical waves $D_1(\mathbf{k})$ and $D_1(\mathbf{k})$ were determined by the boundary conditions imposed on the wave function at the points where the zero-range potentials are centered. The asymptotic behavior of the wave function (0) far from the target determines the amplitude of the elastic scattering of a particle by the pair of identical centers. The total scattering cross section was obtained in [4] using the optical theorem [5].

If the scattering centers in the target are not identical [6] then the boundary conditions imposed on the wave function (0) at the points $\mathbf{R}_1$ and $\mathbf{R}_2$ are described by the following formulas:

$$\psi(\mathbf{r})_{\mathbf{r}\to\mathbf{R}_1} \approx C_1\left[\frac{1}{|\mathbf{r}-\mathbf{R}_1|}+k\cot\delta_1\right];\quad \psi(\mathbf{r})_{\mathbf{r}\to\mathbf{R}_2} \approx C_2\left[\frac{1}{|\mathbf{r}-\mathbf{R}_2|}+k\cot\delta_2\right]. \tag{1}$$

Here $C_1$ and $C_2$ are some constants, $k$ is the particle linear momentum relative to the target, $\delta_1(k)$ and $\delta_2(k)$ are the s-phase of particle scattering on the first and second zero-range potentials of the target, correspondingly. Applying formulas (1) to the wave function (0) and passing to the limits, we obtain the scattering amplitude in a closed form as in [4] but not in the form of an expansion in the partial waves.

In the next section, we derive the basic formulas for calculating the phases of particle scattering on a target formed by a pair of non-identical zero-range potentials. Using the derived formulas, we calculate the phase shifts and cross sections for electron scattering by the model molecules in Section 2. The angular parts of electron continuum wave functions in the field of two centers are explored in Section 3. Section 4 contains conclusion remarks.

## 1. Main formulas

The variational principles formulated in [1] enable one to determine the partial waves and proper phases by direct methods without calculating the closed-form scattering amplitude. According to [1,2], the solution of the scattering problem for a two-center target is obtained by imposing the boundary conditions (1) on the pair of the following partial functions:

$$\psi_\lambda(\mathbf{r}) = \sum_{j=1}^{2} D_j \frac{\sin(k|\mathbf{r}-\mathbf{R}_j|+\eta_\lambda)}{|\mathbf{r}-\mathbf{R}_j|}, \tag{2}$$

where $D_j$ are the unknown coefficients, $\eta_\lambda(k)$ are the phase shifts of the wave function of a particle scattered by the target. As a result, we obtain a system of homogeneous equations, the solution of which determines a set of the phase shifts $\eta_\lambda(k)$.

The system of equations for a two-center target reads

$$\begin{aligned} D_1 k(\cot\eta - \cot\delta_1) + D_2[\operatorname{Im} a \cot\eta + \operatorname{Re} a] = 0 \\ D_1[\operatorname{Im} a \cot\eta + \operatorname{Re} a] + D_2 k(\cot\eta - \cot\delta_2) = 0 \end{aligned} \tag{3}$$

where $a = \exp(ikR)/R$ and $R = |\mathbf{R}_1 - \mathbf{R}_2|$ is the distance between the centers. In the system of equations (3), the number of equations is equal to the number of the unknowns. Therefore, this system has a nontrivial solution if and only if the following matrix equals to zero:

$$\begin{Vmatrix} B_1 & A \\ A & B_2 \end{Vmatrix} = B_1 B_2 - A^2 = 0 . \quad (4)$$

Here $A = \sin(kR+\eta)/R$, $B_1 = k(\cos\eta - \sin\eta\cot\delta_1)$ and $B_2 = k(\cos\eta - \sin\eta\cot\delta_2)$. We rewrite the quadratic equation $B_1 B_2 - A^2 = 0$ with the unknown $x = \cot\eta(k)$ in the form

$$\alpha x^2 + \beta x + \gamma = 0 , \quad (5)$$

by introducing the following notation:

$$\begin{gathered} z = kR, \\ \alpha = (\sin^2 z - z^2) , \\ \beta = [\sin 2z + z^2(\cot\delta_1 + \cot\delta_2)] , \\ \gamma = [\cos^2 z - z^2 \cot\delta_1 \cot\delta_2] . \end{gathered} \quad (6)$$

Solving the quadratic equation (5) we obtain for the first and second phase shifts

$$\cot\eta_0(k) = \frac{-\beta - \sqrt{\beta^2 - 4\alpha\gamma}}{2\alpha}; \qquad \cot\eta_1(k) = \frac{-\beta + \sqrt{\beta^2 - 4\alpha\gamma}}{2\alpha} . \quad (7)$$

Let us see how formulas (7) transform in the case when the powers of the zero-range potentials forming the target become equal. Equating the phase shifts $\delta_1(k)$ and $\delta_2(k)$ and substituting $\delta = \delta_1 = \delta_2$ into equation (7) we obtain the first and second phase shifts for the particle-target scattering [3]

$$\cot\eta_0(k) = \frac{kR\cot\delta - \cos kR}{kR + \sin kR}, \qquad \cot\eta_1(k) = \frac{kR\cot\delta + \cos kR}{kR - \sin kR} . \quad (8)$$

The phase shifts $\eta_\lambda(k)$ in (7) and (8) can be classified by considering their behavior at $k \to 0$ [1,2]. In this limit, the particle wavelength $\lambda\!\!\!^- = 1/k$ is much greater than the target size, and the scattering picture should approach a spherical symmetric one. From these equations we obtain the asymptotic behavior $\eta_0(k \to 0) \sim k$ and $\eta_1(k \to 0) \sim k^3$, correspondingly. Thus, the phase shifts at $k \to 0$ behave similarly to the *s*- and *p*- phase shifts in a spherically symmetric potential. This fact explains the choice of their indexes.

Thus, the general formulas (7) determine the molecular phases of scattering on two-center targets formed by a pair of zero-radius potentials, regardless of whether they are identical or not.

## 2. Numerical calculations of the phase shifts and cross sections

We now apply equations (6-8) to calculation of the scattering phases $\eta_\lambda(k)$ of a slow electron elastically scattered by $C_2$ and CH molecules modeled by a pair of zero-range potentials with the following parameters in equation (6):
$\delta_1(k) = 2\pi - 1.912 \cdot k$ [7] for C atom,
$\delta_2(k) = \pi - 5.72682 \cdot k + 3.62932 \cdot k^2$ [8] for singlet state of H atom,
$R$ = 2.116 au (atomic units) [9] for CH molecule.

The evaluated phases are then used to calculate the average effective cross section $\bar{\sigma}(k)$ for electrons elastically scattered by the target. That is the total cross section integrated over all the angles between the vectors **k** and **k'** (the initial and final wave vectors of the scattered electron, correspondingly) followed by averaging over all directions of the vector **k** in the target reference frame. This cross section is determined by the expression [1,2]

$$\bar{\sigma}(k) = \frac{4\pi}{k^2} \sum_\lambda \sin^2 \eta_\lambda(k) . \qquad (9)$$

Here $\eta_\lambda(k)$ are the phase shifts obtained above. The index $\lambda$ is used to number the phase shifts for nonspherical targets. In the spherically symmetric case, the index $\lambda$ should be replaced by two indices $l$ and $m$ numbering the angular and magnetic quantum numbers, correspondingly. Summation over $m$ leads to the appearance of the factor (2$l$ + 1) in front of sine in Eq. (9). Then summation is conducted over the orbital angular momentum $l$ only. The results of the cross-section calculations for the CH molecule are shown in Fig. 1. For comparison, the curves for partial cross sections for $C_2$ molecule with phase shifts (8) in which $\delta(k) \equiv \delta_1(k)$ are also displayed. As expected, the curves for $C_2$ in Fig.1 coincide with those obtained in [6] with the optical theorem [5].

## 3. Angular parts of molecular continuum wave functions

The potential of our model target is nonspherical. The solution $\psi^+_{\mathbf{k}}(\mathbf{r})$ of the Schrödinger equation with this potential cannot be presented at an arbitrary point of space as an expansion in the spherical harmonics $Y_{lm}(\mathbf{r})$. However, at asymptotically great distances from the target, according to [1,2], the wave function can be presented as an expansion in a set of other orthonormal functions $Z_\lambda(\mathbf{k})$

$$\psi^+_{\mathbf{k}}(\mathbf{r} \to \infty) \approx 4\pi \sum_\lambda R_{k\lambda}(r) Z_\lambda(\mathbf{r}) Z^*_\lambda(\mathbf{k}) \qquad (10)$$

with the radial parts of the wave function determined by the following expression:

$$R_{k\lambda}(r \to \infty) \approx e^{i(\eta_\lambda + \frac{\pi}{2}\omega_\lambda)} \frac{1}{kr} \sin(kr - \frac{\pi}{2}\omega_\lambda + \eta_\lambda) . \qquad (11)$$

The elastic scattering amplitude for a non-spherical target, according to [1,2], is given by the following expression:

$$F(\mathbf{k},\mathbf{k}') = \frac{2\pi}{ik}\sum_{\lambda}(e^{2i\eta_\lambda} - 1)Z_\lambda^*(\mathbf{k})Z_\lambda(\mathbf{k}') . \qquad (12)$$

The cross section $\bar{\sigma}(k)$ averaged over all the directions of momentum of incident electron **k** is connected with the molecular phases $\eta_\lambda(k)$ by the expression (9).

Passing to the limit $r \to \infty$ in the partial functions $\psi_\lambda(r)$ we obtain the following expressions for the asymptotic of the radial parts of the partial waves (2):

$$R_{k0}(r \to \infty) \approx e^{i\eta_0}\frac{1}{kr}\sin(kr + \eta_0) , \qquad (13)$$

$$R_{k1}(r \to \infty) \approx e^{i(\eta_1 - \pi/2)}\frac{1}{kr}\sin(kr - \frac{\pi}{2} + \eta_1) . \qquad (14)$$

The angular parts of the wave functions (10) and (11) are determined by the functions $Z_0(\mathbf{k}')$ and $Z_1(\mathbf{k}')$ [10]

$$Z_0(\mathbf{k}') = \frac{\cos(\mathbf{k}'\cdot\mathbf{R}/2)}{\sqrt{2\pi S_+}}, \qquad Z_1(\mathbf{k}') = \frac{\sin(\mathbf{k}'\cdot\mathbf{R}/2)}{\sqrt{2\pi S_-}} . \qquad (15)$$

Here **R** is the vector along the target axis, $\mathbf{k}' = k\mathbf{r}/r$ is the particle wave vector after the collision, the functions $S_\pm = 1 \pm \sin kR/kR$. The set of the functions $Z_\lambda(\mathbf{k}')$, resembling the spherical harmonics, forms an orthonormal system

$$\int Z_\lambda^*(\mathbf{k})Z_\mu(\mathbf{k})d\Omega_k = \delta_{\lambda\mu} . \qquad (16)$$

Evaluation of the limits of the functions $Z_\lambda(\mathbf{k}')$ at $k \to 0$ in formulas (15) results in the well-known spherical harmonics

$$Z_0(\mathbf{k}')_{k\to 0} \to \frac{1}{\sqrt{4\pi}} \equiv Y_{00}(\mathbf{k}'), \qquad Z_1(\mathbf{k}')_{k\to 0} \to \sqrt{\frac{3}{4\pi}}\cos\vartheta \equiv Y_{10}(\mathbf{k}') . \qquad (17)$$

Here $Y_{lm}(\mathbf{k}') \equiv Y_{lm}(\vartheta_k, \varphi_k)$ ($\theta_k$ and $\varphi_k$ are spherical angles of the vector **k'**). We can follow these limit transitions in Figures 2 and 3. Solid lines in these figures are the limits of the functions $Z_\lambda(\mathbf{k}')$ when $k \to 0$.

Thus, the angular parts of the wave functions of the molecular continuum far from the target do not depend on whether the zero-radius potentials forming the target are identical or not. That is, the formulas (15) are applicable to any two-center targets.

## 4. Conclusions

It is well known that the great advantage of zero-range potentials method is a possibility of obtaining an exact solution of many important and very difficult problems where other methods fail [11]. For many years, this method was used mainly in nuclear physics. See, for example, the work by Brueckner [4] mentioned above. In both our present paper and previous one [3], in addition to the exact closed-form formulas for the scattering amplitude obtained in [4], the exact formulas for the same amplitude were derived in the form of its expansion into a series of partial waves (12). It naturally rises a question: Can these exact formulas be reproduced if we solve our problem with the traditional method of molecular physics?

In molecular physics, the problem of elastic scattering of a slow electron on a diatomic molecule can be considered as an analogue of our problem. But we read in [12] about the wave functions of the molecular continuum: *the continuum electron functions are similar to those for electron-atom scattering*. The asymptotic form of the wave function far from the molecule (beyond the molecular sphere) is, therefore, a sum of a plane wave and spherical waves emitted by the molecular center

$$\psi^{+}_{\mathbf{k}}(\mathbf{r}\to\infty)\approx e^{i\mathbf{k}\cdot\mathbf{r}}+F(\mathbf{k},\mathbf{k}')\frac{e^{ikr}}{r}\,. \qquad (18)$$

Function $F(\mathbf{k},\mathbf{k}')$ here is the elastic scattering amplitude expanded into a series of the spherical functions

$$F(\mathbf{k},\mathbf{k}')=\frac{2\pi}{ik}\sum_{l,m}(e^{2i\eta_l}-1)Y^{*}_{lm}(\mathbf{k})Y_{lm}(\mathbf{k}')\,. \qquad (19)$$

Hence, in molecular physics the solution of the problem of electron scattering by a nonspherical potential is reduced (without any ground by introducing an artificial molecular sphere [13]) to the usual S-matrix method of partial waves for a spherical target. This recipe of building the molecular continuum wave functions is considered as a matter-of-course and, as far as we know, is beyond any doubt. That is, we find that essentially the same problems are solved by different methods. It is hard to imagine that the phase shifts of the wave function of a meson scattered on a deuteron [4] are determined by the matching conditions on the surface of a sphere, surrounding the deuteron, by matching the Schrodinger equation solutions inside and outside the sphere.

The exact formulas for the wave functions of continuum far from two centers targets in the present paper were obtained in explicit form, that include both the radial (13) and (14), and angular parts (15) of functions. It follows from these formulas that no matter how far we move away from the target, the angular parts of the continuum functions $Z_{\lambda}(\mathbf{k}')$ never transform into the spherical functions $Y_{lm}(\mathbf{k}')$. And so we come to the conclusion that the traditional approach based on formula (18) as a result allows one to obtain the scattering phases of a particle on an isolated sphere, rather than on a real two-center target.

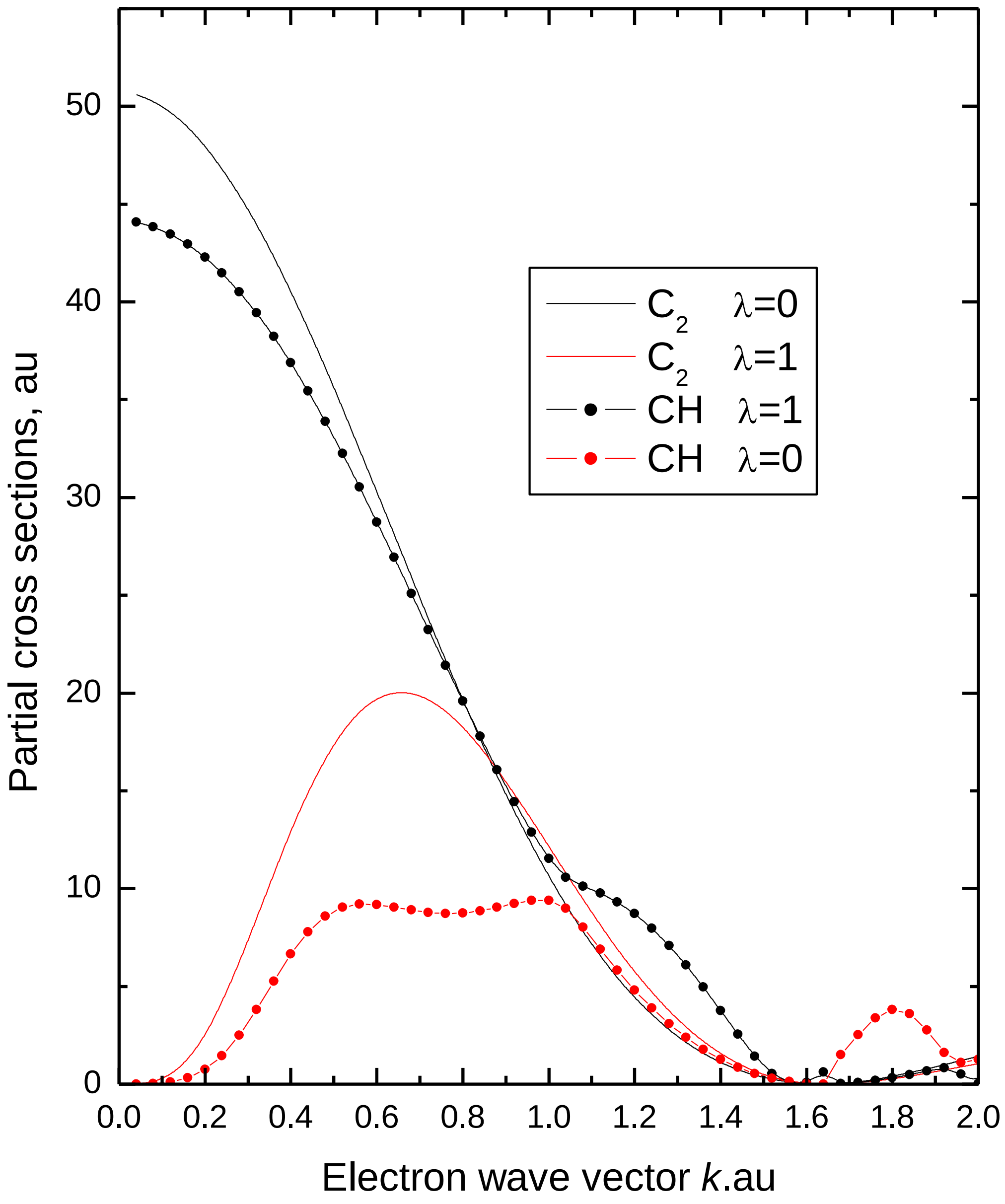

**Figure 1.** Partial averaged cross sections of electron scattering on the molecules CH and $C_2$ versus the electron momentum.

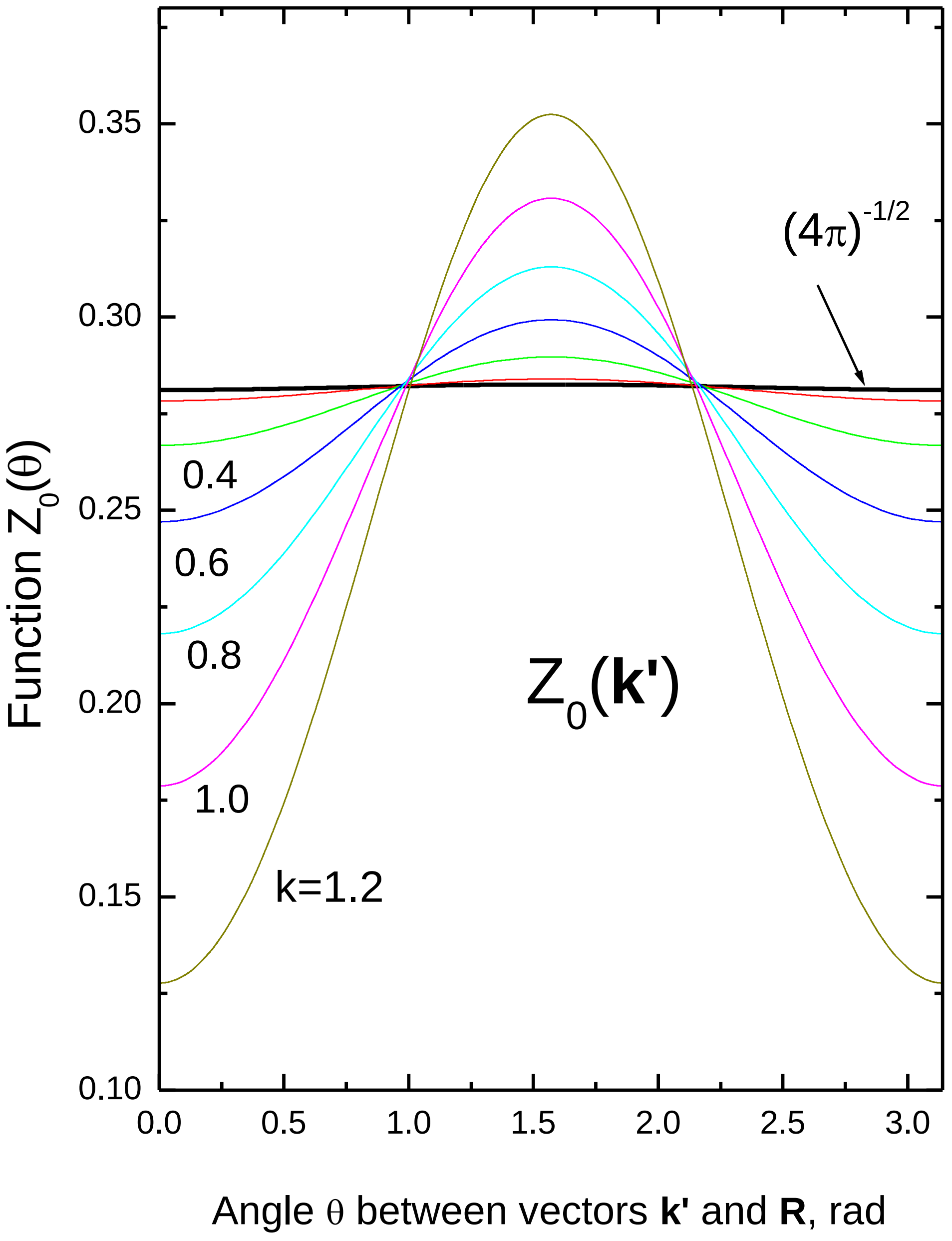

**Figure 2.** Function $Z_0$(**k**) versus the angle $\theta$ between the vectors **k**'= k **r**/r and **R**. Function $Z_0$(**r**) is represented by the same curves, and the angle $\theta$ is the angle between the vectors **r** and **R**

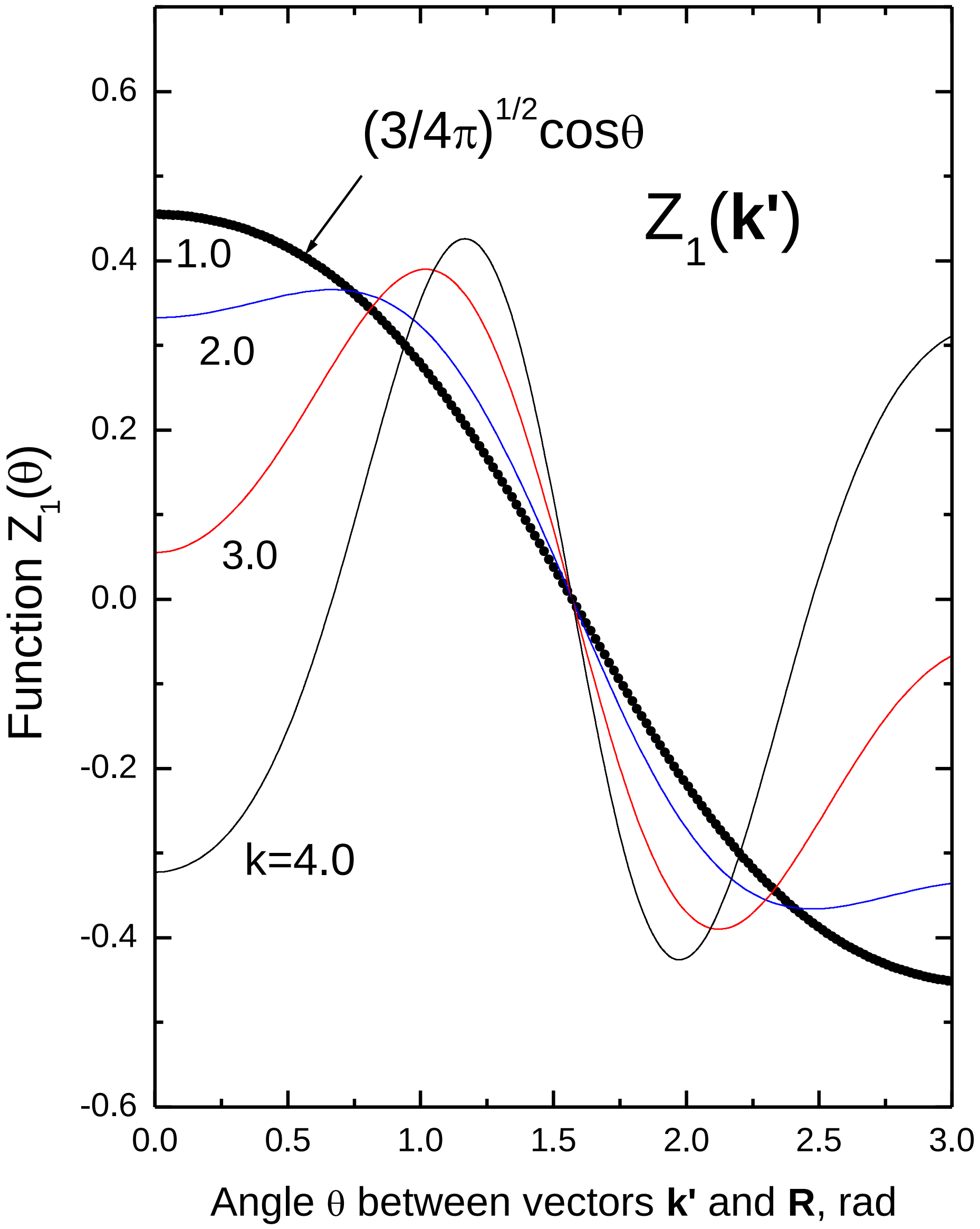

**Figure 3.** Function $Z_1$(**k**') versus the angle $\theta$ between the vectors **k**′ = k**r**/r and **R**. Function $Z_1$(**r**) is represented by the same curves, and the angle $\theta$ is the angle between the vectors **r** and **R.**